\documentclass[final,5p,times,twocolumn]{elsarticle}

\usepackage{amssymb}
\newcommand{\beq}{\begin{equation}}
\newcommand{\eneq}{\end{equation}}
\usepackage{graphicx}
\usepackage{amsmath}
\usepackage{epsfig}
\usepackage{bm}
\input{epsf}
\usepackage{dcolumn}

\usepackage{lineno}

\usepackage{color}

\newcommand{\ket}[1]{\left| #1 \right\rangle}
\newcommand{\bra}[1]{\left\langle #1 \right|}
\newcommand{\proj}[1]{\ket{#1}\bra{#1}}
\newcommand{\Tr}{\mathrm{Tr}}

\journal{Physics Letters A}

\begin{document}

\begin{frontmatter}

\title{Plasmon-assisted quantum control of distant emitters}

\author[ad1]{Cristian E. Susa}
\author[ad1,ad2]{John H. Reina}
\ead{john.reina@correounivalle.edu.co}
\author[add3]{Richard Hildner}
\address[ad1]{Departamento de F\'isica, Universidad del Valle, A.A.
25360, Cali, Colombia}
\address[ad2]{Departamento de \'Optica, Facultad de F\'{\i}sica,
 Universidad Complutense, 28040~Madrid, Spain}
\address[add3]{Experimentalphysik IV, Universit\"at Bayreuth,
Universit\"atsstrasse 30,
95447 Bayreuth, Germany}

\begin{abstract}
We show how to generate and control the correlations in a set of two distant quantum emitters coupled to a one-dimensional dissipative plasmonic waveguide. 
An external laser field enhances the dimer's steady-state correlations and allows an  active control (switching on/off) of nonclassical correlations.
The plasmon-assisted dipolar-interacting qubits 
exhibit persistent correlations, which in turn 
 can be  decoupled and  made to evolve independently from each other.
The setup enables long-distance ($\sim 1\, \mu$m)
qubit control that works for both resonant and detuned emitters.
For suitable emitter initialization, we also show that the quantum correlation is always greater than the classical one.
\end{abstract}


\end{frontmatter}

%\linenumbers

\section{Introduction}
The construction of large quantum networks with controllable long-distance coupling of qubits is one of the key goals in quantum information science~\cite{Ladd}. 
To generate the required qubit-qubit correlations (microwave or optical) photons are typically used. Recently, a novel approach based on single quantum emitters coupled to one-dimensional plasmonic waveguides has been proposed, 
and entanglement of qubits mediated by surface plasmons in such a setup has been explored~\cite{tudela,tudela2}.  Surface plasmons are collective excitations that have become an important physical resource in many applications in physics, chemistry, and materials science~\cite{barnes,mark}. In  particular, the coupling of single emitters to plasmonic structures~\cite{lukin,wrachtrup,hulst} has attracted substantial attention, because it allows the manipulation of the emission properties as well as the enhancement of the interaction between quantum emitters in the vicinity of these structures~\cite{nano,oulton,soren2}. 

Non-local quantum correlations and entanglement  in quantum systems under the action of decoherence effects have been arduously investigated not only because of their  fundamental  physical  implications (see e.g., \cite{zurek,jh}), but also because of their utmost relevance to the development of novel quantum technologies~\cite{Ladd}. The influence of the system-bath coupling over the reduced system's quantum correlations has been studied in  different physical systems such as quantum dots \cite{qdots}, superconducting qubits \cite{supercond},  atoms and photons \cite{atoms}, and biomolecular systems  \cite{jhbio}, to cite just but a few. 
We have recently reported on the distribution of classical and quantum correlations that arise in a bipartite emitter system coupled to 
a plasmonic waveguide \cite{Susa2013}. We have shown that such quantum correlations are more robust through 
the dissipative dynamics than the classical ones, for several experimentally accessible scenarios, and that the emitters collective properties 
that arise from the interaction with the plasmons allow an additional degree of  quantum control on the
correlations dynamics \cite{Susa2013,Susa2}. 

In this Letter, we give a protocol for actively enabling a plasmon-assisted long-distance 
(about~$1\, \mu$m) qubit conditional dynamics between emitters which are externally-driven by a coherent laser field.
%%%%
Such a quantum mechanism works for both  resonant and detuned qubits, and is optimized by laser pumping and by 
tailoring the separation between the emitters. Moreover, from the dimer's conditional entropy, we analytically 
identify specific conditions for which the quantum correlation is always greater than the classical 
correlation~(cf.~\cite{Vedral,lindblad,popescu}). 

In section \ref{theoryI} we introduce the quantifiers of quantum and classical correlations, 
and the corresponding quantum master equation used to describe the emitters' dissipative dynamics is given in section~\ref{theoryII}. 
The quantum control that arises from the 
correlations dynamics is set for: resonant emitters without laser excitation 
(section~\ref{resonant}), resonant emitters
under the action of a coherent laser field (section~\ref{rlaser}), and for emitters with different 
transition energies~(section \ref{detuned}). Concluding remarks are given in section~\ref{concl}. 

\section{Quantum correlations}
\label{theoryI}
The quantum mutual information $
\mathcal{I}(\rho_{AB})=S(\rho_A)+S(\rho_B)-S(\rho_{AB})$
 gives a measure of the total correlation in a  bipartite qubit state $\rho_{AB}$~\cite{popescu,hamieh,Vedral2}, 
and this may be separated into purely quantum---$\mathcal{Q}(\rho_{AB})$ (e.g., via the quantum discord $\mathcal{D}(\rho_{AB})$)~\cite{Vedral2,Zurek,luo2}
and classical correlation---$\mathcal{C}(\rho_{AB})$~\cite{Vedral}: $\mathcal{I}(\rho_{AB})=\mathcal{Q}(\rho_{AB})+\mathcal{C}(\rho_{AB})$. The details of calculation of $\mathcal{C}$ and $\mathcal{Q}$ are left to the appendix. The von Neumann entropy $S(\rho)=-\Tr\rho\log_2\rho$, and 
$\rho_{A(B)}=\Tr_{B(A)}\rho_{AB}$ is the reduced density operator of the partition $A(B)$.
Discord has been linked to the computational speedup in an efficient model of quantum computation~\cite{datta} and has been pinpointed as a valuable resource in quantum information protocols~\cite{lanyon,datta}.

Although Lindblad conjectured that, for any quantum state,  
$\mathcal{C}(\rho)\geq \mathcal{Q}(\rho)$~\cite{popescu,Vedral,lindblad}, we give an entropy condition for which the quantum 
correlation is  always greater than the classical one~\cite{luoJ,Susa2}: since
$\mathcal{I}=\mathcal{Q}+\mathcal{C}$, we ask if  the inequality $\mathcal{I}-2\,\mathcal{C}\geq 0$  is ever met. 
The sought entropy bound reads

\begin{equation}
 2 \mathcal{S}(\rho_{A|\Pi_j^B})-S(\rho_{AB})+S(\rho_{B})-S(\rho_{A})\geq 0\, ,
  \label{encond}
\end{equation}
where $\mathcal{S}(\rho_{A|\Pi_j^B})=\mathrm{min}_{\{\Pi_j^B\}}\{\sum_jp_jS(\rho_{A|\Pi_j^B})\}$, 
$S(\rho_{A|\Pi_j^B})$ is the entropy associated to the density matrix
of subsystem $A$ after the measure. Equation~\eqref{encond} is indeed satisfied, at all times, for suitable resonant emitters.

To distinguish the quantum correlations that arise for entangled states and separable states, we quantify the 
emitters'  entanglement via the entanglement of formation $\mathcal{E}_{\mathcal{F}}$. For two-qubit systems,~$\mathcal{E}_{\mathcal{F}}(\rho_{AB})=\varepsilon\big([1+\sqrt{1-C^2(\rho_{AB})}]/2\big)$,   
where
$
  \varepsilon(r)=-r\log_2r-(1-r)\log_2(1-r) 
$ 
denotes the binary entropy function, and the concurrence 
$C(\rho_{AB})={\mathrm{max}}\{0, \lambda_1- \lambda_2- \lambda_3-\lambda_4\},$ 
where the $\lambda_i$'s are,  in decreasing order, the eigenvalues of the matrix
$\sqrt{\rho_{AB} \tilde\rho_{AB}}$;  $\tilde\rho_{AB} =
(\sigma_y\otimes\sigma_y) \bar\rho_{AB} (\sigma_y\otimes\sigma_y)$,  $\bar\rho_{AB}$
is the elementwise complex conjugate of $\rho$, and $\sigma_y$ is the Pauli matrix \cite{wootter}. 
This entanglement metric is of entropic character  and can be compared on the same grounds with the discord \cite{Susa2013}.

\section{Plasmon-emitter master equation}
\label{theoryII}
We consider a pair of distant emitters that act as a dipole-dipole 
two-qubit system via the interaction with the plasmon modes in a metallic nanostructure. The total Hamiltonian of the system plus environment 
can be written as ($\hbar$ is the reduced Planck's constant):
\begin{eqnarray}
	\hat{H}&=&\hat{H}_{S} + \hat{H}_{E} + \hat{H}_{int}%\nonumber\\
	=\sum_{i=1}^2\frac{1}{2}\hbar\omega_i\hat{\sigma}_z^{(i)} +\\ \nonumber
	&&+\int d^3\boldsymbol{r}\int_0^{\infty}d\omega\hbar\omega \hat{f}_{\omega}^{\dagger}(\boldsymbol{r})\hat{f}_{\omega}(\boldsymbol{r})
	-\sum_{i=1}^2\hat{\mu}_i\cdot\hat{E}(\boldsymbol{r}_i) \;,
\label{TOH}
\end{eqnarray}
where the first two terms denote the free energy of the two-qubit system and 
the free energy of the electromagnetic field represented by a bosonic bath, respectively;  $\hat{\sigma}_z^{(i)}$ is the $z$ Pauli matrix associated to the $i$-th emitter of transition 
frequency $\omega_i$, and $\hat{f}_{\omega}(\boldsymbol{r})$, and 
$\hat{f}_{\omega}^{\dagger}(\boldsymbol{r})$ are the bosonic excitation 
operators of the quantized electromagnetic field with the usual commutation relations 
$[\hat{f}_{\omega}(\boldsymbol{r}),\hat{f}_{\omega'}^{\dagger}(\boldsymbol{r}')]=\delta(\omega~-~\omega')\delta(\boldsymbol{r}~-~\boldsymbol{r}')$, and 
$[\hat{f}_{\omega}(\boldsymbol{r}),\hat{f}_{\omega'}(\boldsymbol{r}')]=[\hat{f}_{\omega}^{\dagger}(\boldsymbol{r}),\hat{f}_{\omega'}^{\dagger}(\boldsymbol{r}')]=0$.
The last term represents the interaction between the dipole operator 
$\hat{\mu}=\boldsymbol{\mu_i}\hat{\sigma}_+^{(i)} + \boldsymbol{\mu_i}^*\hat{\sigma}_-^{(i)}$ and the quantized field operator 
$\hat{E}(\boldsymbol{r})=\hat{E}^+(\boldsymbol{r}) +\mathrm{H.c.}$, where 
$\sigma^{(i)}_{+}=\ket{1_{i}}\bra{0_{i}}$ ($\sigma^{(i)}_{-}=\ket{0_{i}}\bra{1_{i}}$) are the raising 
(lowering) Pauli operators acting on  emitter $i$ ($\ket{0}$ and $\ket{1}$ denote the ground and first excited state which represent the qubit computational basis), and $\hat{E}^+(\boldsymbol{r})=\int_0^{\infty}d\omega\hat{E}(\boldsymbol{r},\omega)$ is the positive frequency part,

\begin{equation}
\nonumber
	\hat{E}(\boldsymbol{r},\omega)= \mathrm{i}\sqrt{\frac{\hbar}{\pi\epsilon_0}}\frac{\omega^2}{c^2}\int d^3\boldsymbol{r}'\sqrt{\epsilon''(\boldsymbol{r}',\omega)}\boldsymbol{G}(\boldsymbol{r},\boldsymbol{r}',\omega)\hat{f}_{\omega}(\boldsymbol{r}') .
\label{efield}
\end{equation}
The Green's tensor $\boldsymbol{G}(\boldsymbol{r},\boldsymbol{r}',\omega)$ satisfies the
Maxwell-Helmholtz wave equation, supports the electromagnetic 
interaction from $\boldsymbol{r}'$ to $\boldsymbol{r}$, and contains all the information about the coherent and incoherent properties of the system, i.e., about the dipole-dipole shift and the 
different channels of radiation through the vacuum (far field) and through the metal. 
$\epsilon''(\boldsymbol{r},\omega)$ is the imaginary part of the 
electric permittivity of the metal, and this is considered as a constant value corresponding to the 
permittivity of the silver at the operational wavelength described here, and $\epsilon_0$ is the permittivity of the vacuum.

The dynamics of the total system  (emitters plus electromagnetic field) can be derived in the 
Schr\"odinger picture from the 
Liouville-von Neumann equation 
$
	\dot{\rho}_{S-E}=-\frac{\mathrm{i}}{\hbar}\left[\hat{H},\rho_{S-E}\right] ,
$ 
which, in the interaction picture (denoted by the superscript $I$), can be written as the following integro-differential equation 
\begin{eqnarray}
	\dot{\rho}^I_{S-E}(t)&=&-\frac{\mathrm{i}}{\hbar}\left[\hat{H}^I_{int}(t),\rho^I_{S-E}(0)\right]-\nonumber\\
	&&-\frac{1}{\hbar^2}\int_0^tdt'\left[\hat{H}^I_{int}(t),\left[\hat{H}^I_{int}(t'),\rho^I_{S-E}(t')\right]\right] ,
	\label{intpict}
\end{eqnarray}
where 
$\rho^I_{S-E}(t)=e^{\mathrm{i}(\hat{H}_{S}+\hat{H}_{E})t/\hbar} \rho_{S-E} e^{-\mathrm{i}(\hat{H}_{S}+\hat{H}_{E})t/\hbar}$, and 
$\hat{H}^I_{int}(t)=e^{\mathrm{i}(\hat{H}_{S}+\hat{H}_{E})t/\hbar} \hat{H}_{int} e^{-\mathrm{i}(\hat{H}_{S}+\hat{H}_{E})t/\hbar}$.~Here, we consider  atom-like (small) emitters with an operational wavelength in the optical frequency regime (we use $\lambda_{0}=640$~nm) coupled to a broadband plasmonic waveguide. We perform the Born-Markov  approximation ($\rho^I_{S-E}(t')=\rho^I(t')\otimes\rho_{E}(0)$, and  $\rho^I(t')\rightarrow\rho^I(t)$) in order to solve Eq.~\eqref{intpict} due to the weak coupling between the emitters and the plasmonic electromagnetic field~\cite{tudela,tudela2,soren2,hummer}. 
%Born ($\rho^I_{sys-env}(t')=\rho^I(t')\otimes\rho_{E}(0)$) and Markov %($\rho^I(t')\rightarrow\rho^I(t)$) 
%approximation in Eq.~\eqref{intpict}. 
Additionally, we avoid rapidly oscillating terms much higher than $\omega_i$ by 
applying the rotating wave approximation (RWA) to the interaction part of the Hamiltonian in Eq. \eqref{TOH}, such that
$\hat{H}_{int}=-\boldsymbol{\mu}_1\sigma_+^{(1)}\hat{E}^{\dagger}(\boldsymbol{r}_1)-\boldsymbol{\mu}_2\sigma_+^{(2)}\hat{E}^{\dagger}(\boldsymbol{r}_2)+\mathrm{H.c.}$.

Tracing out Eq. \eqref{intpict} over  the environment degrees of freedom, 
and going  back to the Schr\"odinger picture we can, under the above assumptions, 
describe the dimer's dissipative dynamics by means of the following quantum master equation~\cite{soren2,Susa2,jh04}

\begin{equation}
{\dot\rho}= \frac{i}{\hbar}\left[\rho,\hat{H}_{\mathrm{eff}}\right]
 -\sum_{i,j=1}^{2}\frac{\Gamma _{ij}}{2}\left( {\rho}
      \sigma^{(i)}_{+}\sigma^{(j)}_{-}+\sigma^{(i)}_{+}\sigma^{(j)}_{-}{\rho}
      -2\sigma^{(i)}_{-}{\rho} \sigma^{(j)}_{+}\right),
\label{master}
\end{equation}
where the effective dimer's Hamiltonian $\hat{H}_{\mathrm{eff}}=\hat{H}_{S}+\hat{H}_{12}+\hat{H}_L$ 
contains the coherent dipole-dipole shift 
$\hat{H}_{12}=\frac{1}{2}\hbar V(\sigma_x^{(1)}\otimes\sigma_x^{(2)}+\sigma_y^{(1)}\otimes\sigma_y^{(2)})$, and the laser-qubit interaction $\hat{H}_L$ (see section \ref{rlaser}). The strength 
of the effective dipole-dipole interaction is calculated as:

\begin{equation}
	  V=\frac{1}{\pi\epsilon_0 c^{2} \hbar}\mathcal{P}
  \int_0^{\infty}d\omega\frac{\omega^2\mathrm{Im}[\boldsymbol{\mu}^\ast_1 
    \boldsymbol{G} (\omega,\mathbf{r}_1,\mathbf{r}_2)\boldsymbol{\mu}_2]}
  {\omega - \omega_0} ,
\label{cohpart}
\end{equation}
where $\omega_0=(\omega_1+\omega_2)/2$ and $\mathcal{P}$ denotes the principal part of 
the integral.  The second term of Eq.~\eqref{master} accounts for the dimer's  incoherent effects 
that arise from radiation into the vacuum, dissipation through losses in the metal, and the excitation of the surface plasmons that propagate 
on the metallic surface~\cite{tudela2,tudela3}. If the losses in the metal are negligible, the two remaining dissipative channels to be accounted for are captured by 

\begin{equation}
	\Gamma_{ij}=\frac{2\omega_0^2}{\epsilon_0c^2\hbar}\mathrm{Im}
  [\boldsymbol{\mu}^\ast_i \boldsymbol{G} (\omega_0,\mathbf{r}_i,\mathbf{r}_j)\boldsymbol{\mu}_j],
\label{incohpart}
\end{equation}
where $\Gamma_{11}=\Gamma_{22}\equiv\Gamma$, and 
$\Gamma_{12}=\Gamma^{\ast}_{21}\equiv \gamma$ are the individual and collective 
spontaneous emission rates, respectively~\cite{jh04}, and  $i,j=1,2$.
The emitters-plasmon dissipative coupling  renders new dipole-dipole strengths $V$ and collective 
damping rates $\gamma$. If we consider that the emitters' dipoles are equally 
oriented and are located at the same distance to the surface of the metal, 
and that the emitters'  most relevant  dissipative contribution is due to the plasmon 
modes, an exact calculation of $V$ (Eq.~\eqref{cohpart}) and  $\gamma$ 
(Eq.~\eqref{incohpart}) follows from the Green's 
functions of the plasmon excitations~\cite{tudela,tudela2,nano,soren2}:

 \begin{eqnarray}
    V=\frac{\Gamma}{2}\beta e^{-\frac{\lambda_{pl}}{2 L} \zeta}\sin{(2 \pi \zeta)} ;   
 \;   \; 
   \gamma=\Gamma\beta e^{-\frac{\lambda_{pl}}{2 L} \zeta}\cos{(2 \pi \zeta)}, \;   \;   \;   \;   \;
   \label{collective}
 \end{eqnarray}
where  the wavelength of plasmons
$\lambda_{pl}\equiv 2\pi / k_{pl}$, 
$k_{pl}$ is the plasmon wave vector,
$\zeta\equiv d/\lambda_{pl}$, $d$ gives the distance between emitters, and 
$L$ is the propagation length of the plasmons.

Equations~\eqref{collective} are valid for high $\beta$-factors 
($\beta\equiv \Gamma^{\mathrm{guided}}/\Gamma$, $\Gamma=\Gamma^{\mathrm{vac}}+\Gamma^{\mathrm{guided}}$), 
which 
means that the dimer's  main  dissipative mechanism is due to the 
propagating plasmons on the metal surface, and hence the fraction of emitted radiation  by the plasmonic propagating mode $\beta\mapsto 1$. This factor depends on the vertical distance between 
the emitters and the metallic surface. Here, we consider geometries like $\Lambda$-wedge
or $V$-groove as our plasmonic waveguides, for which the plasmonic excitations have been shown to
substantially enhance the interqubit interaction~\cite{nano,oulton}, thus validating  
Eqs.~\eqref{collective}. In particular,  for  the $V$-groove 
channel, $\beta$-factors $\sim0.91$ (and higher)  are expected for  an emitter-metal vertical separation above the bottom of the $V$-channel waveguide of about $160$~nm,  and a depth of the channel  of $140$~nm~\cite{tudela,tudela2,nano}. For such a channel depth, we note that shorter emitter-metal distances ($<80$~nm) imply a reduction in the $\beta$-factor since the losses into the metal become appreciable. On the other hand, larger emitter-metal distances ($> 180$~nm), also diminish 
 the $\beta$-factor since radiation into the 
vacuum becomes the major decay contribution \cite{tudela2}. 
In addition, the `plasmonic approximation' given by Eqs.~\eqref{collective} 
works well for emitter-emitter separations larger than $\sim
\lambda_{\mathrm{pl}}/4$~\cite{tudela2,soren2}, and in this case
$\lvert \gamma \rvert\leq\Gamma$ and $\lvert V \rvert\leq\frac{\Gamma}{2}$. Below such distance, the vacuum 
electromagnetic fluctuations become more significant and modify the collective 
parameters in equations \eqref{cohpart} and \eqref{incohpart}.

A general two-qubit state can  be written as 
$
\epsilon=\frac{1}{4}\big(I_{4\times4}+\vec{a}\vec{\sigma}\otimes I_{2\times2}+I_{2\times2}\otimes \vec{b}\vec{\sigma}+\sum_{i=1}^{3}h_{i}\sigma_{i}\otimes\sigma_{i}\big)
$,  where $\vec{a}=(a_1,a_2,a_3), \vec{b}=(b_1,b_2,b_3) \in \mathbb{R}^3$, and $h_{i}\in \mathbb{R}$~\cite{luo2,fano}. 
The initial density matrix used in our calculations adopts an $X$-like structure 
with $\vec{a}=(0,0,a_3), \vec{b}=(0,0,b_3)$, and $h_1=h_2=\eta$:
\begin{eqnarray}\label{rhosim}
 \rho_{AB}(0) & = & \nonumber 
\left(
\begin{array}{cccc}
\frac{1+c_++h_3}{4}  & 0 & 0 & 0 \\
0 & \frac{1+c_--h_3}{4} & \frac{\eta}{2} & 0 \\
0 & \frac{\eta}{2} & \frac{1-c_--h_3}{4} & 0 \\
0 & 0& 0 & \frac{1-c_++h_3}{4}
\end{array}\right) ,
\end{eqnarray}
where $c_{\pm}=a_3\pm b_3$, and  $a_3$, $b_3$, $\eta$, and $h_3$ are chosen 
such that $\rho_{AB}(0)$ is a well-defined density matrix \cite{fano}.
\begin{figure}[t]
  \centering
  \includegraphics[width=5.5cm]{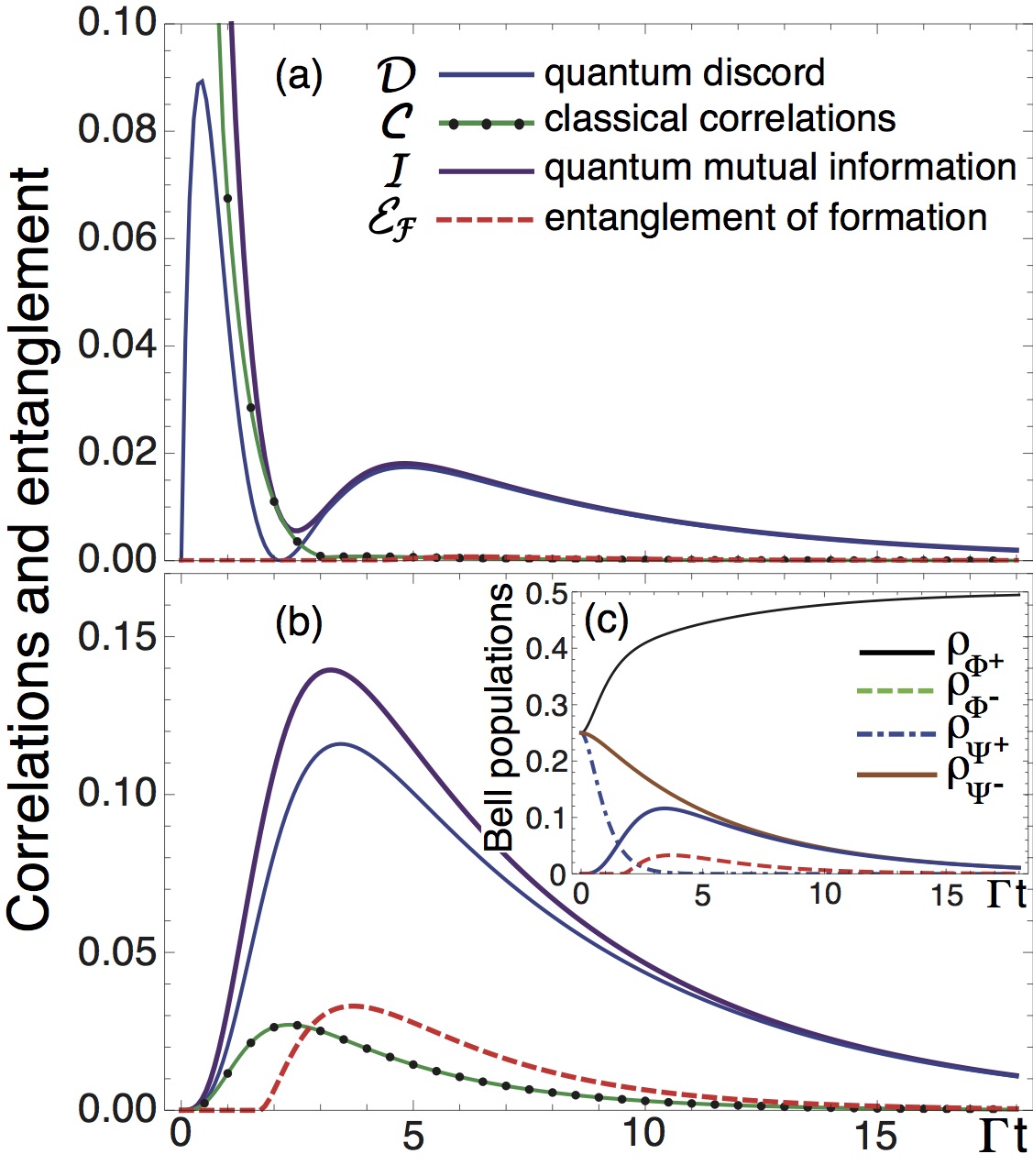}
  \caption{Dynamics of correlations for initial states  (a) 
  $\rho_{\mathcal{C}}$, 
and (b) 
$\rho_{MM}$. 
We employ optimal plasmonic waveguide parameters~\cite{tudela}: $L=2\,\mu$m, $\beta=0.94$, $\lambda_{pl}=542$~nm 
(operational wavelength $\lambda_{0}=640$~nm). The emitters are 
separated a distance $d=\lambda_{pl}$ ($\zeta=1$). 
(c) Bell states populations for the case (b).}
  \label{fig1}
\end{figure}

\section{Resonant molecules without laser pumping} 
\label{resonant}
For the above initial state, and identical emitters with transition frequencies 
$\omega_1=\omega_2\equiv\omega_0$,, Eq.~\eqref{master}  admits an analytical solution in the absence of optical driving. The non-trivial density matrix elements read 
 $a(t)\equiv\rho_{11}= 1-b_{+}(t)-b_{-}(t)-f(t)$,  where $f(t)\equiv\rho_{44}=\frac{1}{4} p e^{-2 \Gamma t}$, with $p=1+h_{3}-c_+$, and

\begin{eqnarray}
\label{rhot}
&& b_{\pm}(t)  \equiv 
\left\{
\begin{array}{c}
\rho_{22}^+   \\
\rho_{33}^- \\
\end{array}\right\}= \\
&&\frac{e^{-\Gamma t }}{4 \gamma_{-}^2} \Big[-p \gamma_{+}^2e^{-\Gamma t }- \left(c_+\gamma_{+}^2-2 \left(\Gamma^2+h_{3}\gamma^2\right) \right) \cosh{(\gamma t )-} \nonumber \\ 
&& 2 \left(p \gamma \Gamma +\eta \gamma_{-}^2\right) \sinh{(\gamma t )}\pm c_- \gamma_{-}^2 \cos{(2  V t)}\Big] ,\nonumber \\
&& z(t)\equiv \rho_{23}= \nonumber \\ 
&& \frac{e^{-\Gamma t }}{4 \gamma_{-}^2} \Big[-2p\gamma\Gamma e^{-\Gamma t }+2\left(\eta\gamma_{-}^2+
p\gamma\Gamma\right)\cosh{(\gamma t )}+ \nonumber \\ 
&& \left(c_+\gamma_{+}^2-2\left(\Gamma^2+h_{3}\gamma^2\right)\right)\sinh{(\gamma t )}+
\mathrm{i}c_-\gamma_{-}^2\sin{(2  V t)}\Big] , \nonumber 
\end{eqnarray}
and  $z^*(t)\equiv\rho_{32}$, with $\gamma_{\pm}^2=\Gamma^2\pm\gamma^2$. 
Importantly, the  initial states $\rho(0)$  contain the eigenstates of the system's Hamiltonian 
($\hat{H}_{S}+\hat{H}_{12}$):  $\ket{00}$, $\ket{\Psi^{\pm}}=\frac{1}{\sqrt{2}}(\ket{01}\pm\ket{10})$, $\ket{11}$, 
which allows for their experimental generation, control, and read-out \cite{vahid,richard}.

From  Eqs.~\eqref{rhot} we see that for $c_-=0$ 
the density matrix becomes  
independent of the dipolar interaction $V$ ($b_+~=~b_-~=~b$) and, as such, the quantum correlation dynamics is only due to 
the collective damping $\gamma$. Such correlations persist for long times and  for  a large emitters' separation 
($d=\lambda_{pl}; \zeta=1$), as shown in Fig.~\ref{fig1}. 
For the plasmon waveguide we used realistic parameters $L=2\,\mu$m, 
$\beta=0.94$, and $\lambda_{pl}\approx542$~nm, that follows from the dispersion 
relation reported in Ref.~\cite{tudela}, based on an operational wavelength 
$\lambda_{0}=640$~nm. These parameters can be achieved with e.g. 
plasmonic wedges ($\Lambda$) and V-groove waveguides.
Although we have chosen $\zeta=1$, which makes $V=0$, our results are more general and can be extrapolated to other $\zeta$-values because the considered initial states evolve completely independently of $V$. 
\begin{figure*}[t]
  \centering
  \includegraphics[width=12cm]{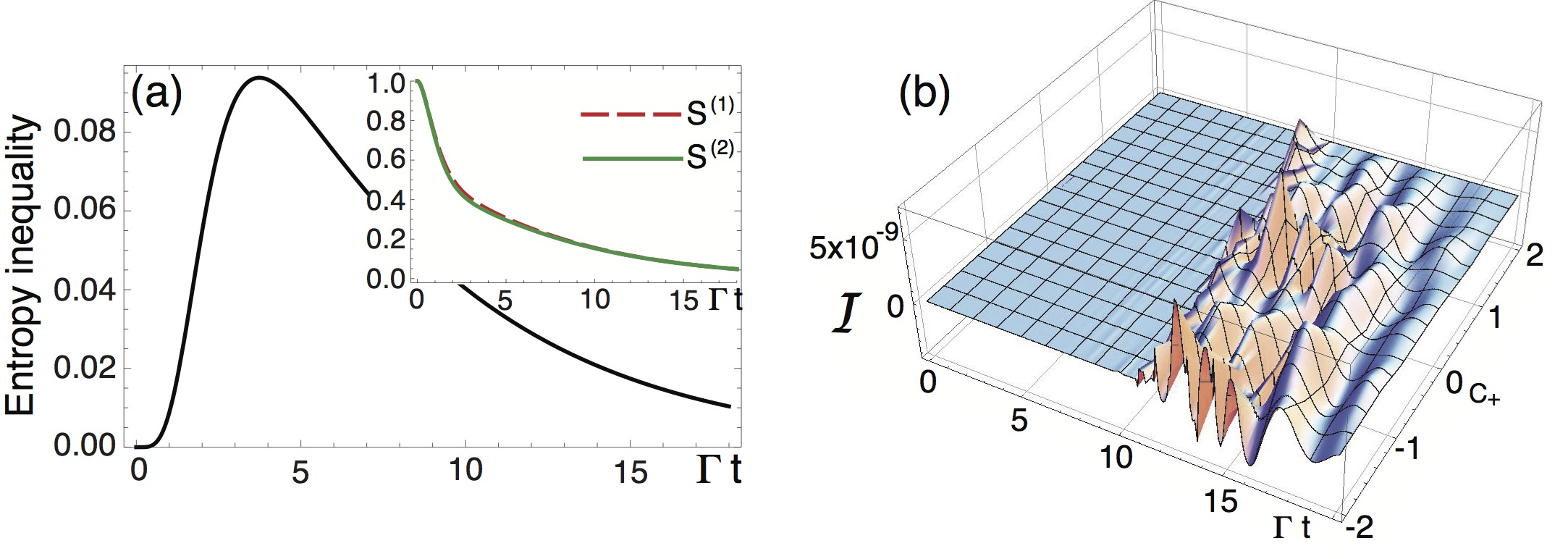}
  \caption{
  (a) Left-hand side of inequality Eq.~(\ref{encond}), for the case shown 
  in Fig.~\ref{fig1}(b). The inset shows the evolution of the quantities $S^{(1)}$ (dashed red curve) and $S^{(2)}$ (solid green curve), whose minima give the conditional entropy. (b) Numerical simulation of the quantum mutual information 
for the set of initial states $\Xi:=\{c_-=0$, $\eta=0$, $h_3=c_+^2/4\}$ in Eqs.~\eqref{rhosim}. The same result 
  is obtained for the other correlations.}
  \label{fig2}
\end{figure*}

Figure \ref{fig1}(a)  shows the correlations dynamics for the classically correlated initial state 
$\rho_{\mathcal{C}}=\frac{1}{2}\proj{00}+\frac{1}{2}\proj{11}$ 
for which $\mathcal{C}=\mathcal{I}=1$. Further, in Fig.~\ref{fig1}(b) we show how the correlations are 
increased for a completely uncorrelated initial state 
$\rho_{MM}=\mathrm{diag}\{\frac{1}{4},\frac{1}{4},\frac{1}{4},\frac{1}{4}\}$. 
Note that while the quantum 
discord increases from $t=0$, the emitters follow entanglement sudden birth \cite{yu09} for $t\sim4/\Gamma$ (Fig.~\ref{fig1}(a)), and $t\sim2/\Gamma$ (Fig.~\ref{fig1}(b)). 
For $t>10/\Gamma$ Fig.~\ref{fig1}(a) shows that the system becomes disentangled, 
 quantum discord, however, equals
$\mathcal{I}$ and the nonclassical correlation represents the total correlation in the dimer (i.e., $\mathcal{C}=0$).~Figure~\ref{fig1}(c) plots the population dynamics of the Bell basis states $\rho_{\Psi^\pm}~\equiv~\ket{\Psi^\pm}\bra{\Psi^\pm} $ and $\rho_{\Phi^\pm}$, $\ket{\Phi^{\pm}}=\frac{1}{\sqrt{2}}(\ket{00}\pm\ket{11})$, as well as the time evolution of discord 
and entanglement
for $\rho(0)=\frac{1}{4}(\rho_{\Psi^+}+\rho_{\Psi^-}+\rho_{\Phi^+}+\rho_{\Phi^-})$ [Fig.~\ref{fig1}(b)]:  
although the  entanglement of formation ($\mathcal{E}_{\mathcal{F}}$) eventually vanishes, the dynamics of $\mathcal{D}$ quickly approaches that of the antisymmetric Bell state $\rho_{\Psi^-}$, which decays at the slowest rate $\Gamma-\gamma$; this is due to the natural mixture of the Bell populations which spontaneously create quantum correlations beyond entanglement, here captured by  discord.
Figure~\ref{fig1} thus shows that, for suitable (unentangled)  initialization of the emitters, discord is the most robust and persistent nonclassical correlation available.

An important point to note is that for the initial state 
$\rho_{MM}$ 
($c_-=0$) the quantum correlation is {\it always} greater than the classical correlation. Under these initial conditions the inequality given by Eq.~\eqref{encond} is satisfied; they have  an  associated conditional
entropy  $\mathcal{S}(\rho_{A|\Pi_j^B})=\mathrm{min}_{\{\Pi_j^B\}}(S^{(1)},S^{(2)})$, with 
\begin{eqnarray}
S^{(1)}&=&-a\mathrm{log}_2\frac{a}{a+b}-b\mathrm{log}_2\frac{b}{a+b}  -\\
&&b\mathrm{log}_2\frac{b}{b+f}-f\mathrm{log}_2\frac{f}{b+f} , \nonumber \\ 
S^{(2)}&=&-\frac{1}{2}(1-\xi)\mathrm{log}_2\frac{1}{2}(1-\xi) \nonumber -\frac{1}{2}(1+\xi)\mathrm{log}_2\frac{1}{2}(1+\xi),
\label{twoS}
\end{eqnarray}
where $\xi^2=(a-f)^2+|z|^2$,  the marginals $S(\rho_A)=S(\rho_B)$, and we arrive at
$2 \mathcal{S}(\rho_{A|\Pi_j^B})-S(\rho_{AB})\geq 0$. A direct calculation of $S^{(1)}$ and $S^{(2)}$ then demonstrates that the quantum correlation is greater than the classical one, as can be seen in Fig.~\ref{fig2}(a), where we have plotted the entropy bound (Eq.~(\ref{encond})). 
 This behaviour  has also been demonstrated for other emitters' preparations~\cite{Susa2013}.

If we now choose the emitters' mutual separation such that $\zeta_n=\frac{1}{4}(2n+1)$, $n$ 
integer, the damping $\gamma$ vanishes according to Eqs.~\eqref{collective}. As the emitters' dynamics is independent of $V$, the density matrix elements read
\begin{eqnarray}
\nonumber 
f(x)&=&\frac{1}{4}px^2; \; \; \; \;  z(x)=\frac{1}{2}\eta x, \\
b_{\pm}(x)&=& \frac{1}{4}x\left[2\left(1-\frac{c_+}{2}\right)-px\right], \\
\nonumber 
a(x) &=& 1+\frac{1}{4}x\left[px+4\left(\frac{c_+}{2}-1\right)\right],
\end{eqnarray}
where $x\equiv e^{- \Gamma t}$. For this 
family of initial states (which includes the eigenstates of $\hat{H}_{\mathrm{eff}}$ in absence 
of the coherent external laser), the dynamics of 
all  correlations has an asymptotic decay following the time parameter $x$, in contrast to the 
scenario shown in Fig.~\ref{fig1} for which $\gamma$ is non-zero. If we add the  conditions $\eta=0$ and $h_3=c_+^2/4$, the density  matrix adopts a diagonal 
structure without quantum correlations. This also allows us to show that  
$\rho_{AB}=\rho_{A}\otimes\rho_{B}$,  and hence $\mathcal{C}=0$. Since  $\mathcal{I}(\rho_{AB})$ 
gives the relative entropy between the matrices $\rho_{AB}\,\mathrm{and}\,\rho_{A}\otimes\rho_{B}$, 
it can also be seen that $\mathcal{I}(\rho_{AB})=0$, and hence  there are neither quantum nor 
classical correlations for this time evolution. 
Figure~\ref{fig2}(b) shows the numerical simulation for the quantum mutual information as a function 
of the parameter $c_+$ (the fluctuations around $\sim 10^{-9}$ are due to numerical errors and 
do not imply $\mathcal{I}\neq0$). 
Thus the system evolves through a path of product states even though 
$V\neq 0$. 
This provides a useful method for controlling the dynamics of 
correlations in two-qubit systems.

\section{Resonant molecules under laser pumping}
\label{rlaser}
\begin{figure*}[t]
  \centering
  \includegraphics[width=11cm]{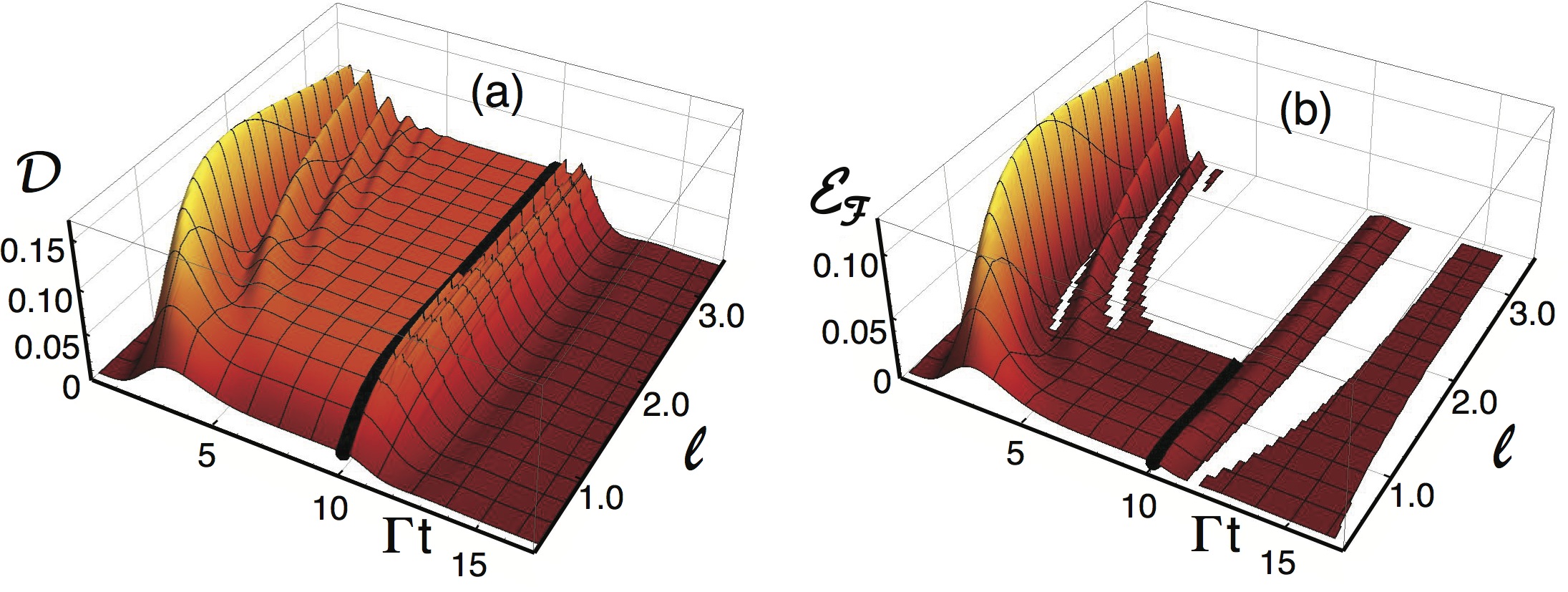}
  \caption{ 
  (a) Quantum discord and (b) entanglement of formation  as functions of the laser field amplitude $\ell$,
for resonant qubits $\omega_L=\omega$, 
and initial (ground) state $\ket{00}$. Waveguide parameters are as in Fig.~\ref{fig1};
$d=3\lambda_{pl}/4 $, so $\gamma=0$. The black solid curve indicates the time at which the laser is turned off.
}
  \label{LD1}
\end{figure*}

Illumination with a  laser field, in conjunction with the emitters-plasmon coupling, 
allows a reestablishment of the  correlations, which were lost due to the $V$-independence of the matrix elements. The light-emitter interaction is given by 
$\hat{H}_L=\sum_i \hbar \ell_{i}(\sigma^{(i)}_{-}e^{\mathrm{i}\omega_{L}t}+\sigma^{(i)}_{+}e^{-\mathrm{i}\omega_{L}t})$, 
with $\hbar\ell_{i}=-\hat{\mu}_i\cdot \hat{E}^{(L)}_i$, $i=1,2$~\cite{jh04}.

To illustrate this mechanism, 
we consider two individual molecules initially in their ground states,
with resonance wavelength 
$\lambda_0$ in order  to reach the optimal waveguide parameters of 
Fig.~\ref{fig1}.  We now, however, consider a shorter emitters' separation: $\zeta_1=3/4$,  such that $\gamma=0$. 
The chosen values for the waveguide parameters at $\lambda_0$ 
simulate structures such as plasmonic wedges 
($\Lambda$) and V-groove waveguides, which have been theoretically demonstrated 
 to exhibit the greatest plasmon mode propagation 
\cite{tudela,nano,oulton}.
Figure~\ref{LD1} depicts  the evolution of the quantum discord (graph (a)) and 
the entanglement of formation (graph (b)) as functions of time (in $1/\Gamma$ units), and 
laser intensity, under the choice 
$\ell_{1}\equiv \ell\gg\ell_{2}$. As an illustration, we fix $\ell_{2}=0$ and $\ell\in[0.4 \Gamma,3.6 \Gamma]$ such  
that the effect of the coherent light over the emitter 1 is much stronger than that over
emitter 2 (the latter being negligible). This particular choice is meant to highlight our main results and does not imply any restriction on the validity of the mechanism here presented or loss of generality for other choices of laser intensity.

By turning on the  laser  field (at $t = 0$) to illuminate 
emitter 1 with an intensity $\ell$, 
we reach a time regime for which  the correlations dynamics stabilizes ($\mathcal{D}\sim0.06$), 
as conservatively indicated by the thick black curve in Fig.~\ref{LD1} at $\Gamma t\sim 10$.
If we then turn off the laser field at $\Gamma t = 10$, 
the emitters  decay again to the ground state well before $\Gamma t\sim 15$. The conditional dynamics between the laser field  and the dipolar interaction favours such a behaviour. 
We point out that if, in the same scenario, the dipole-dipole interaction were zero 
(non-interacting qubits or emitters in vacuum with interqubit separation 
$d=3\lambda/4$), the correlations would then equal zero even if the laser were turned on (not shown).  We have also checked that the scheme works for other input states 
which satisfy the initial condition 
$\Xi:=\{c_-=0$, $\eta=0$, $h_3=c_+^2/4\}$ 
(Fig.~\ref{fig2}(b)). 
This emphasizes the relevance of the  plasmon-mediated interaction between the emitters, because this scheme lets us switch on/off the collective parameters $\gamma$ and $V$ (Eqs. \ref{collective}).

Figure~\ref{LD1} also demonstrates that
the quantum discord becomes the `most relevant' correlation at high laser intensity: $\mathcal{E}_{\mathcal{F}}$ rapidly decays
at high intensities and, at about 
$\ell \sim  1.5\, \Gamma$ (see the `white zone' in graph (b)) the emitters have already experienced collapses, revivals, and early stage 
disentanglement~\cite{yu09}. In contrast, the discord reaches nonzero stationary 
values (Fig.~\ref{LD1}(a)), indicating the presence of  nonclassical
correlations different to entanglement. 

\section{Detuned molecules} 
\label{detuned}
\begin{figure}[h]
  \centering
  \includegraphics[width=5.5cm]{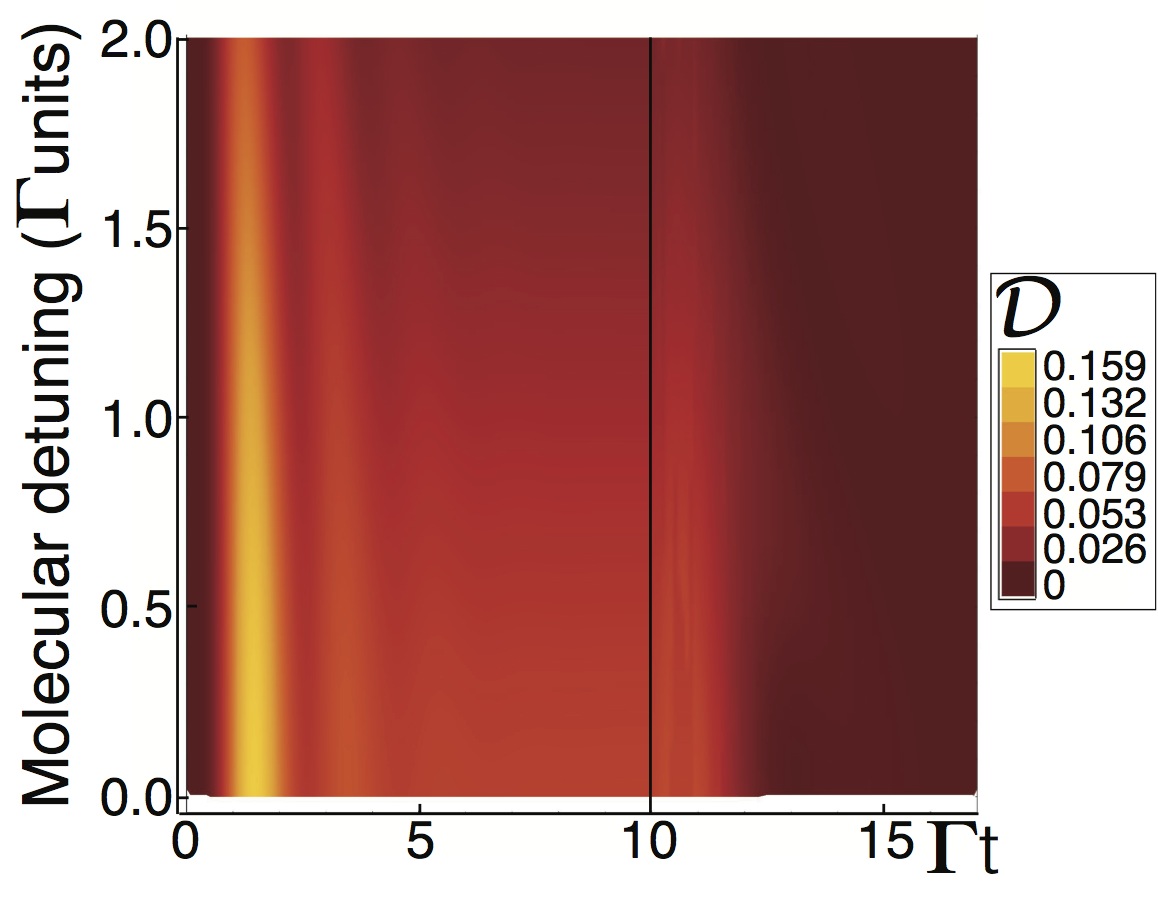}
  \caption{Density plot of the switching mechanism for the quantum discord as function of time and  molecular detuning $\delta$ under laser pumping with amplitude $\ell=1.5\Gamma$,
  and laser frequency detuning $\Delta=\omega_L-\omega_0$. 
  Initial state and waveguide parameters are as in Fig.~\ref{LD1}. The black vertical line indicates the 
  time at which the laser is turned off.}
  \label{LD2}
\end{figure}

As yet, we have assumed identical emitters to show the  mechanism for switching correlations.  We notice, however, that this quantum control also works in a 
more general physical setup of two molecules with different transition 
frequencies, $\omega_1\neq\omega_2$, i.e. with a frequency detuning $\delta=\omega_1-\omega_2$. In fact, this situation is encountered in any experiment on molecular systems~\cite{vahid,richard0}.

In this case, the eigenstates of the emitters Hamiltonian are: 
$\ket{00}$, 
$\ket{11}$, and two 
new entangled states $\ket{\phi^+}=\alpha_1\ket{01}+\alpha_2\ket{10}$, and 
$\ket{\phi^-}=\alpha_2\ket{01}-\alpha_1\ket{10}$, with  $\alpha_1=\sqrt{V^2/(\kappa^2+V^2)}$,
$\alpha_2=\sqrt{\kappa^2/(\kappa^2+V^2)}$, and 
$\kappa=\delta/2+\sqrt{V^2+(\delta/2)^2}$. The new transition frequencies of these 
eigenstates are, respectively: $
\omega_{\pm}=\omega_0 \pm \sqrt{V^2+(\delta/2)^2}$.
The inclusion of a molecular detuning does not allow for a general analytical solution of the master equation (\ref{master}).
If we 
assume, however, that the emitters have interqubit separation $\zeta_n$, and are initialized 
by  $c_-=0$,  
in the absence of laser excitation 
we obtain

%
%\begin*{\textwith}
\begin{eqnarray}
\label{SolnI}
z(t)&=&2\eta\mathcal{G} \left[4V^2 + \delta^2 \cos{\left(\Upsilon t\right)} 
+ \mathrm{i} \delta \Upsilon \sin{\left(\Upsilon t\right)} \right] , \\ \nonumber
 b_{\pm}(t)&=&\mathcal{G} 
    \Big[\left((2 - c_+)-p e^{-\Gamma t}\right) \Upsilon^2\pm 4\eta \left(\cos{\left(\Upsilon t\right)}-1\right)  \delta  V\Big] ,
\end{eqnarray}
%\end*{\textwith}
%
where $a(t)$ and $f(t)$ are given as in Eqs.~\eqref{rhot}, 
$\mathcal{G}=\mathrm{exp}(-\Gamma t)/4\Upsilon^2$, 
and $\Upsilon^2=4 V^2 + \delta^2$. In contrast to the previous case of identical molecules, 
this evolution is not independent of $V$ so the correlations may exhibit oscillatory behaviour 
due to the terms in 
$z(t)$. As we are interested in a rapid switching of 
correlations, we apply the complete configuration $\Xi$ to the initial condition such that the 
matrix element $z(t)$ goes to zero, and the factor $\Upsilon^2$  simplifies in $b_{\pm}(t)$. The reduction of  Eqs. (\ref{SolnI}) shows that, under the parameters choice  \{$\Xi$, $\zeta_n$\}, the 
same result as for identical, resonant molecules is recovered: all correlations remain zero.

To understand the influence of the molecular energy mismatch under laser pumping, we 
plot the quantum discord in terms of $\delta$ in Fig.~\ref{LD2}. 
We illuminate emitter 1 with a laser intensity $\ell=1.5\Gamma$,
and take into account the eigenenergies of the intermediate 
states to introduce the laser frequency detuning 
$\Delta:=\omega_L-\omega_0=\sqrt{V^2+(\delta/2)^2}$. This choice makes the dynamical action of the two entangled eigenstates effective, so both the quantum discord and 
the entanglement can be enhanced. The value reached by the discord slightly decreases
as the molecular detuning increases from $0$ to $2.0\, \Gamma$, because  the  emitters start to be uncoupled from each other for $\delta \gg V$.

The fact that, for our initial conditions, the interqubit coupling $V$ is maximized for the collective decay rate $\gamma=0$ may be exploited to perform quantum logic gates. 
For the interqubit dipolar coupling here considered, $\hat{H}_{12}$, one can define an optimal entangling  gate 
 for the dipole-dipole interaction; this, plus local unitaries, give a universal gate-set~\cite{jh04}. 
The plasmon-assisted dipolar  interaction $V$, for e.g. $\zeta_1=3/4$, in combination with an applied laser field, would then allow the realization of  conditional quantum dynamics. As we have demonstrated, this works for molecules separated by distances of at least 0.4 $\mu$m. 

\section{Concluding remarks} 
\label{concl}
We stress that our approach can be tested in the laboratory with 
present day technology.
The plasmonic structure is modelled with parameters that 
can be achieved
with e.g.~either~$\Lambda$ or V-groove channel waveguides~\cite{boz}. 
Although we note that our calculations are based on ideal plasmonic structures, 
the basic phenomena that we have theoretically predicted should still be observable 
in an experiment using real structures.
Perylene dyes are suitable single emitters because they are highly photostable, 
very bright and well characterized at the single-molecule level~\cite{richard0}. 
The deterministic placement of single emitters on plasmonic structures~\cite{hulst}
can be achieved by lithographic methods or chemical surface functionalization. 
For the initialization and coherent control of this plasmon-coupled dimer system, 
(high resolution) laser addressing combined with time-correlated single-photon 
counting techniques are readily available~\cite{vahid,richard0,richard}.
These allow the quantification of the molecules' dipolar coupling strength, the decay 
rates, and photon emission required in the generation of  the quantum states and 
the correlations here described.

It is worth pointing out that for 
other waveguide geometries, the computed $\beta$-factor can reach values well below those used here. For example, for a cylindrical waveguide the highest predicted $\beta$-factor is $\sim 0.6$~\cite{tudela2,nano}. In such a case, the parameter window for which both the dimer to waveguide surface 
distance, and the 
emitter-emitter separation fall within the plasmonic approximation Equations~\eqref{collective}  becomes  limited, and  the losses into the metal, as well  as the radiation into vacuum play 
more significant roles and hence an evaluation of the full Green's tensor is required. This said,  the main effect on the results here reported would be a reduction in the degree of quantum correlations in the dimer system, and  the protocol for long-distance quantum control here described would 
continue to work.

In summary, we have demonstrated 
a mechanism for controlling 
and switching
nonclassical correlations between qubits. Our proposal is based on a set of distant quantum emitters, that are coupled via the interaction with the plasmon modes of a waveguide and driven by an external laser field. We  have shown that discord  plays a key role in the proposed  emitters' conditional quantum dynamics protocol. 
We have also given a suitable qubit preparation for which the emitters' quantum correlation is always greater than the classical one. Our results could be exploited for quantum computing and communication at long interqubit distances, with the potential for long-range quantum circuitry integration.

\section*{Acknowledgements}
C.E.S. thanks the Colombian Administrative Department of Science, Technology and Innovation (Colciencias) for a fellowship.
J.H.R. gratefully acknowledges Universidad del Valle for  partial funding under grant CI 7930, and for a leave of absence. We thank L. L. S\'anchez-Soto for fruitful discussions.
R.H. acknowledges the DFG under grant GRK 1640.

\appendix
\section{Classical and quantum correlations}

The classical correlation is the
maximum extractable classical information from a partition, say  $A$, when a set 
of positive operator valued measures (POVMs)
has been performed on the other partition ($B$)~\cite{Vedral}.
The measurement of classical correlation, 
as introduced in Ref.~\cite{Vedral}, is given by:

\begin{equation}
  \label{CC}
  \mathcal{C}(\rho_{AB}) = \sup_{\{\Pi_j^B\}} \Big[
    S(\rho_{A})-\sum_j p_j S(\rho_{A|\Pi_j^B})\Big]  ,
\end{equation}
where  
$S(\rho_{A|\Pi_j^B})$ is the entropy associated to the density matrix
of subsystem $A$ after the measure. Such correlation must be
non-increasing, and invariant under local unitary operations, and
$\mathcal{C}(\rho_{AB}) = 0$ if and only if $\rho_{AB}= \rho_A\otimes
\rho_B$.

The set of POVMs that maximize the classical correlation is a complete unidimensional 
measurement $\{\Pi_j^B\}$~\cite{hamieh}. 
If $\{\ket{0},\ket{1}\}$ defines the basis states for the qubit 
$B$, the projectors can be written as $\Pi_j^B=\mathrm{\bf 1}\otimes
\proj{j}$, $j=a,b$, where $\ket{a}= \cos{\theta}\ket{0}+e^{\rm
  i\phi}\sin{\theta}\ket{1}$, $\ket{b}= e^{- i\phi}
\sin{\theta}\ket{0}-\cos{\theta}\ket{1}$, and the optimization is
carried out over angles $\theta$ and $\phi$. The measure $\mathcal{C}$
is antisymmetric by definition, and, without loss of generality, we
take the qubit $B$ to be the one measured.

Following the definition for $\mathcal{C}(\rho_{AB})$, a simple way to
introduce the total quantum correlation in a composite bipartite system is
$\mathcal{D}(\rho_{AB})=\mathcal{I}(\rho_{AB})-\mathcal{C}(\rho_{AB})$. In terms of 
the von Neumann entropies, the quantum correlation, which coincides
with the definition for the quantum discord given in Ref.~\cite{Zurek},
reads

\begin{equation}
  \label{D}
  \mathcal{D}(\rho_{AB})= S(\rho_B)-S(\rho_{AB}) +
  \inf_{\{\Pi_j^B\}} \sum_jp_j S(\rho_{A|\Pi_j^B}) \, .
\end{equation}
For pure states, $\mathcal{D}=S(\rho_B)$, and $D=0$ if and only if  the system is purely
classically correlated.

\end{document}